\documentclass[submission]{dmtcs}

\usepackage{graphicx}
\usepackage[latin1]{inputenc}
\usepackage{color}
\usepackage{amssymb}
\usepackage{verbatim}
\usepackage{xspace}
\usepackage{tikz}
\usepackage{amsmath}

\usepackage{ntheorem}

\newtheorem{thm}{Theorem}[section]
\newtheorem{cor}[thm]{Corollary}
\newtheorem{lem}[thm]{Lemma}
\newtheorem{prop}[thm]{Proposition}
\newtheorem{nota}[thm]{Notation}
\newtheorem{defn}[thm]{Definition}
\numberwithin{equation}{section}

\theoremstyle{plain}
\theorembodyfont{}
\newtheorem{rem}{Remark}

\newenvironment{myproof}[1][Proof.]{\begin{trivlist}
\item[\hskip \labelsep {\bfseries #1}] \leavevmode\par}{\hfill $\Box$\end{trivlist}}

\newcommand{\balpha}{\boldsymbol{\alpha}}
\newcommand{\Def}[1]{{\bf #1}}

\newcommand{\ws}{|\textbf{W}_m|}
\newcommand{\bigO}[1]{\mathcal{O}(#1)}
\newcommand{\Em}{[1,\ws]}
\newcommand{\Comment}[1]{{\color{red}#1}}
\renewcommand{\Comment}[1]{}
\newcommand{\Ab}{{\sf{A}}}
\newcommand{\Cb}{{\sf{C}}}
\newcommand{\Gb}{{\sf{G}}}
\newcommand{\Ub}{{\sf{U}}}

\newcommand{\Hyp}{\Lambda}
\newcommand{\subsubsubsection}[1]{\vspace{.2em}\par\noindent {\it #1}}
\newcommand{\Hypo}[1]{{{\sf  H}$_{\text{\sf\bfseries #1}}$}}

\begin{document}

\title{The weighted words collector}%

\author{J\'er\'emie du Boisberranger\addressmark{1} \and
Dani\`ele Gardy\addressmark{1} \and Yann Ponty\addressmark{2}\thanks{Email: \email{yann.ponty@lix.polytechnique.fr}}}
\address{\addressmark{1} Universit\'e de Versailles, PRISM/UMR 8144, Versailles, France\\
CNRS/Ecole Polytechnique/INRIA AMIB, LIX/UMR 7161 X-CNRS, Palaiseau, France}%

\received{2012-04}
\revised{\today}
\accepted{tomorrow}

\maketitle
\begin{abstract}
We consider the word collector problem, i.e. the expected number of calls to a random weighted generator before all the words of a given length in a language are generated.
The originality of this instance of the non-uniform coupon collector lies in the, potentially large, multiplicity of the words/coupons of a given probability/composition.
We obtain a general theorem that gives an asymptotic equivalent for the expected waiting time of a general version of the Coupon Collector.
This theorem is especially well-suited for classes of coupons featuring high multiplicities.
Its application to a given language essentially necessitates knowledge on the number of words of a given composition/probability.
We illustrate the application of our theorem, in a step-by-step fashion, on four exemplary languages, whose analyses reveal a large diversity of asymptotic waiting times,
generally expressible as $\kappa\cdot m^p \cdot (\log m)^{q}\cdot (\log \log m)^{r}$, for $m$ the number of words, and $p, q, r$ some positive real numbers.
\end{abstract}

\section{Introduction}

The choice of a suitable random model for the input instances of an algorithm is critical for its analysis. In an attempt to capture non-uniform distributions naturally arising in real-life data, Denise \emph{et al}~\cite{Denise2010} studied \Def{weighted languages}, a natural generalization of context-free languages~\cite{flajoletcalculus} where atomic weights are associated to each letter. The weight of a word is then simply the product of its letters' own weight. This naturally induces a probability distribution over the class of words of a given length $n$, where the probability of any given word is proportional to its weight. Aside from arguably being the simplest non-uniform generalization of combinatorial classes, such distributions naturally arise in statistical physics (Boltzmann partition function), with direct applications in algorithm design (Monte-Carlo Markov Chains) and bioinformatics~\cite{McCaskill1990}. Random generation algorithms were also proposed for these distributions~\cite{Denise2010}, leading to an efficient multidimensional generalization of Boltzmann sampling~\cite{Bodini2010}.

These distributions, and their associated random generation algorithms, can also be found in bioinformatics, where RNA folding has been one of the leading problems of the past three decades. Given an RNA sequence of length $n$, composed of four types of nucleotides (\Ab, \Cb, \Gb{} or \Ub), the goal is to predict the {secondary structure}, a non-crossing subset of experimentally-determined base-pairs (hydrogen bonds). This coarse-grain representation of the 3D conformation of RNA molecules has been extensively studied from a combinatorial perspective~\cite{waterman78,Vauchaussade85}. A statistical sampling approach proposed by Ding and Lawrence~\cite{DiLa03} is one of the leading methods for tackling this problem. At the core of this method, one makes repeated calls to a \Def{random generation algorithm}, which draws secondary structures with probability proportional to their Boltzmann factor. Unfortunately, such a redundancy is arguably uninformative when the probability of each conformation can be exactly and efficiently estimated after each generation.
One can thus interpret this redundancy as a degradation of the algorithm performance, and analyze the expected time-complexity of generating $k$ \emph{distinct} conformations. In the worst-case scenario, the targeted number $k$ of secondary structures is the total number of secondary structures. Since energy-weighted secondary structures are in bijection with weighted \emph{peakless}-Motzkin words, then the worst-case/average-case (resp. on $k$ and $n$ the length) complexity of the algorithm is exactly the waiting-time of completing the class of weighted Motzkin words of length $n$.

Generalizing on this question, the central problem addressed by this article is that of the \Def{Weighted Words Collector}: Given a formal language and a word length $n$, how many calls to a weighted generation algorithm must be made before all the words of length $n$ are obtained?
This problem is clearly a weighted instance of the ubiquitous \Def{Coupon Collector problem} which, given a finite collection $C_m$ of $m$ items produced by a random source, studies the expected waiting time $E[C_m]$ of the full collection $C_m$, i.e. the expected number of generations before each item in $C_m$ is present in the generated set. This problem naturally arises in a large variety of contexts, including the analysis of database~\cite{Berenbrink2009} and network~\cite{Gardy2002} probabilistic algorithms.
In the specific context of weighted languages, the two main specificities are the non-uniform nature of the weighted distribution and
the potentially large multiplicity of coupons.

In the uniform distribution, either probabilistic or combinatorial arguments can be used to establish that $E[C_m] = m\cdot\mathcal{H}(m)\in \Theta(m\log m)$, where $\mathcal{H}(m)=\sum_{i\ge 1} 1/i$ is the $m$-th harmonic number.
For general distributions,where the $i$-th object is generated with probability~$p_i$, Flajolet, Gardy and Thimonier~\cite{FlaGarThi92} gave a general expression for the waiting time of the full collection:
\begin{equation}
  E[C_m] = \int_0^{\infty}\left(1-\prod_{i=1}^{m}\left(1-e^{-p_i t}\right)\right)dt.\label{eq:couponcollector}
\end{equation}
However, specializing this formula for a given probability distribution seldom leads to spectacular simplifications, and the derivation of asymptotic estimates for parameterized families of items usually remains challenging. To overcome this limitation, many efforts have focused on providing closed-form approximations~\cite{Berenbrink2009}, asymptotic equivalents~\cite{Boneh1996,Neal2008} and algorithms for computing the waiting time over non-uniform distributions of diverse degrees of generality.
Weighted distributions over languages can be seen as highly specialized non-uniform coupon collections, whose major specificity is that many items may share the same probability or, in other words, some probability may appear with \Def{large multiplicity}. Unfortunately, previous results either fail to apply to classes of coupons of high multiplicity, lead to bounds on the asymptotic behavior that are not tight~\cite{Gardy2010}, or require extensive \emph{a priori} knowledge on the distribution, motivating further studies in the context of languages.

Intuitively, the waiting time of a non-uniform instance of the Coupon Collector problem is dominated by the generation of a subset
composed of the least probable items. Indeed, some subset of items can be so improbable that it is typically fully obtained only after
all the other items in the collection are generated. In such cases, a lower bound on the waiting time can be obtained by
isolating the subset and analyzing its waiting-time as a uniform coupon collector problem.
However, deciding which subset to study can be rather challenging, as the waiting time usually arises as a subtle tradeoff between the probability and the multiplicity. In the case of weighted languages, the presence of coupons having, simultaneously, large multiplicities and equally large discrepancy in their probabilities gives rises to a rich variety of asymptotic behaviors, and calls for a sophisticated -- arguably technically involved -- analysis.

After a brief introduction, this extended abstract states, in Section~\ref{sec:mainthm}, a general theorem for weighted families of coupons. More precisely, Theorem~\ref{mainThm} relates the asymptotic behavior of a general Weighted Coupon Collector Problem to the multiplicity and weight of the $i$-th class of coupons. Section~\ref{sec:litterature} compares the scope of the theorem with previous works addressing a similar problem. Section~\ref{sec:application} develops a methodology to ease the verification of the conditions of Theorem~\ref{mainThm} in the case of context-free languages, and applies it on illustrative examples. Finally, we conclude in Section~\ref{sec:conclusion} by summarizing the contribution and describing future developments.

\section{A general theorem for coupons of large multiplicities}
\label{sec:mainthm}

\subsection{Definitions and notations}
Given a sequence $\textbf{w}=\{w_i\}_{i=1}^m$ of positive numbers, or \Def{weights},  associated with a collection $C_m$ of items,
one defines a \Def{weighted probability distribution} $\{p_i\}_{i=1}^m$ over $C_m$ as:
\begin{align*}
	p_i &= \frac{w_i}{\mu(m)}, \forall\ i\leq m &\text{where}&& \mu(m)&=\sum\limits_{i=1}^{m}w_i.
\end{align*}
In this work, we are interested in distributions with high multiplicity, in the sense that multiple items may share the same weight/probability.
Let us then introduce $\textbf{W}_m = \{W_{m,i}\}_{i}$ the increasingly-ordered, finite, sequence of all \Def{distinct weights} in $\textbf{w}$.
Furthermore, for each $i\in \Em$, let us denote by $M_{m,i}$ the \Def{multiplicity} of the weight $W_{m,i}$, i.e. the number of occurrences of $W_{m,i}$ in $\textbf{w}$.
We observe that:
\begin{align*}	
	m &= \sum\limits_{i=1}^{\ws}M_{m,i}& & \text{and} & \mu(m) &= \sum\limits_{i=1}^{\ws}M_{m,i}\cdot W_{m,i}.
\end{align*}

\subsection{Main result}
We describe a first-order asymptotical expression for the expected time of the full collection, assuming a large number $m$ of items.
Accessible weights $\textbf{W}_m$ and their multiplicities $\textbf{M}_m$ may in principle vary for different values of $m$, leading, in the extreme case,
to the absence of a limit expression for the waiting time. Therefore we restrict the scope of our main theorem to distributions that obey three, essentially
technical, conditions.
\begin{itemize}
\item[\Hypo{1} -] The number $m$ of coupons and the weight rank $i$ may interact only in a simple way within the multiplicity of a given weight. Thus we require that:
\begin{itemize}
\item There exists functions $f_1 ,\ldots,f_p $, $g_1,\ldots,g_p$, $h$ and $H$, such that
		\begin{align*}
			M_{m,i} &\underset{m\rightarrow\infty}{\sim} \frac{e^{\sum\limits_{j=1}^{p}f_j(i)g_j(m)}}{h(i)},&\text{and}&&
			M_{m,i} &\leq \frac{e^{\sum\limits_{j=1}^{p}f_j(i)g_j(m)}}{H(i)},\; \forall m\geq 1, \forall i\leq \ws.
		\end{align*}
\item The functions $f_1$ and $g_1$ must effectively determine the growth of $M_{m,i}$, therefore one requires that: $f_1$ is positive and non-zero everywhere, $g_j(m) = o(g_1(m)), \forall j\in[2,p]$, and $g_1(m)\rightarrow+\infty$.
\item Finally, $\sum\limits_{i\in \Em}\frac{1}{H(i)}$ must converge, to prevent $H$ from capturing the growth of $M_{m,i}$.
\end{itemize}
\item[\Hypo{2} -] Similarly, we restrict the possible interactions of the weight rank $i$ and the number $m$ of items within the
$i$-th weight $W_{m,i}$, by requiring the existence of functions $\nu (i) >0$ and $\omega (m) >0$ such that
\begin{align*}
W_{m,i}&\geq \nu (i)\cdot\omega(m),\; \forall m\ge 1,\forall i \ge 1,
\end{align*}
and such that any weight at rank $i$, beyond some value of $m$, remains constant:
\begin{align*}
 \forall k>0, \exists m_k>0 \text{ such that }W_{m,i}&= \nu(i)\cdot\omega(m),\; \forall m\geq m_k, \forall i\leq k.
\end{align*}
\item[\Hypo{3} -] The multiplicity $M_{m,i}$ must not grow too quickly in comparison with the weight $W_{m,i}$.
More precisely, if $\ws\underset{m\rightarrow\infty}{\rightarrow}\infty$, then one must have
		\begin{equation*}
	\lim\limits_{i\rightarrow\infty}\frac{\nu(i)}{f_j(i)} = +\infty, \forall j\leq p.
		\end{equation*}
\end{itemize}
The conditions are sufficient (yet not always necessary) to obtain the asymptotic behavior of the waiting time, and hold for a large class of weighted languages.
\begin{thm}\label{mainThm}
	Assume that, for all $m>0$, the weights $\textbf{W}_m$ and multiplicities $\textbf{M}_m$ of the coupon collection satisfy the conditions \Hypo{1}, \Hypo{2}, and \Hypo{3}.
 Then, as $m\rightarrow\infty$, one has
		\begin{equation}
			E[C_m] = t^\ast(F,\nu)\cdot G(m)\cdot\frac{\mu(m)}{\omega(m)}\cdot (1+o(1)),
		\end{equation}
where:
\begin{itemize}
\item $\mu(m)$ is the total weight of all coupons;
\item $F\equiv f_1$ and $G\equiv g_1$, defined in \Hypo{1}, drive the leading term of the growth of $M_{m,i}$ as $m\to\infty$;
\item $\omega(m)\cdot\nu(1)$ is the smallest weight within the collection of cardinality $m$ (see \Hypo{2});
\item $t^\ast(F,\nu)$ is the largest value of $t$ such that there exists $x\in\mathbb{N}$ such that $F(x)-t\cdot \nu(x)>0$.
\end{itemize}
\end{thm}

\begin{myproof}[Sketch of proof.]
We give here a brief description on our proof, whose details can be found in Appendix~\ref{sec:appendix}.
\begin{figure}\label{figPsi}
  \input{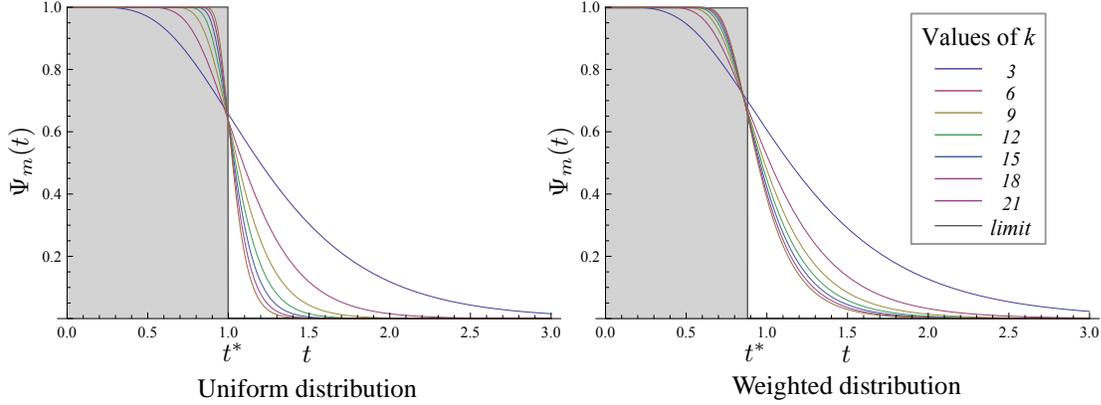}
  \caption{Plots of the $\Psi_m(t)$ functions as appearing for a uniform (Left) and weighted (Right, $\pi(a)/\pi(b)=2/3$) distribution over the $(a+b)^*$ language. We consider $m=2^k$ coupons/words, for several values of  $k\in\{3,6,9,12,15,18,21\}$. The convergence of $\Psi_m(t)$ towards a step function when $m\to\infty$ featuring a transition at $t^{\ast}$ ($t^{\ast}=1$ in the uniform distribution and $t^{\ast}=8/9$ in the weighted one) is crucial to our approach. }
\end{figure}

Applying a substitution $u \frac{\omega(m)}{\mu_{m}\sum\limits_{j=1}^{p}g_j(m)}\rightarrow t$ to Equation~\ref{eq:couponcollector} gives
	\begin{align*}
		E[C_m] &= \frac{\mu_{m}}{\omega(m)}\sum\limits_{j=1}^{p}g_j(m)\int_{0}^{\infty} \Psi_m(t)\ dt & &\text{where }& \Psi_m(t) & := \left[1-\prod\limits_{i=1}^{\ws}\left(1-e^{-t \frac{W_{m,i}}{\omega(m)}\sum\limits_{j=1}^{p}g_j(m)}\right)^{M_{m,i}}\right].
	\end{align*}

Focusing on this expression, one shows that the integral of $\Psi_m(t)$ converges towards some constant. Indeed, numerical computations, as illustrated by Figure~\ref{figPsi}, suggest that $\Psi_m$ converges toward a step function when $m\to\infty$. This can be rigorously proved under the conditions \Hypo{1}, \Hypo{2} and \Hypo{3}, and the integral from 0 to $t^\ast(f_1,\nu)$ converges to $t^\ast(f_1,\nu)$, while the remaining integral converges to $0$.
\end{myproof}

\begin{rem}
	Theorem~\ref{mainThm} applies in the special case of the uniform distribution. Indeed, considering the weight collection is $\{w_i\}_{i=1}^{m}=\{1\}_{i=1}^m$, one has $p_i=1/m$ and $\mu_m=m$. The set of weights is then reduced to the singleton $\textbf{W}_m=(1)$, which has multiplicity $M_{m,1}=m=e^{\log m}$.
\begin{itemize}
\item \Hypo{1} is satisfied upon taking
\begin{align*}F(i)&:=f_1(i)=1, &G(m)&:=g_1(m)=\log m,& h(1)&=1 &\text{and}&& H(1)&=1, \end{align*}
noticing that
$\sum_{1\leq i\leq \ws}1/H(i)=1/H(1)$, which obviously converges.
\item Since one has $W_{m,1}=1$, then \Hypo{2} is satisfied with $\nu(1)=1$ and $\omega(m)=1$.
\item Since $\textbf{W}_m$ is finite, the limit condition of \Hypo{3} does not need to be verified.
\end{itemize}
	One then easily verifies that $t^\ast(f_1,\nu)=t^\ast(1,1)=1$, and applying Theorem~\ref{mainThm} unsurprisingly gives $E[C_m]\sim m\log m$,
which is the well-known asymptotics of the uniform coupon collector.
\end{rem}

\section{Comparison with existing results}
\label{sec:litterature}
Let us compare the scope of our result with previous work on the subject; we remind the reader that $m$ is the number of coupons/words, which typically grows exponentially along with $n$ the length of words.
Given the rich literature dealing with variations on the Coupon Collector problem (e.g. waiting time of first occurrence of a $k$-duplicated collection~\cite{Adler2003}), we will restrict our comparison to three results that are representative of the main approaches used to tackle the problem.

\subsection*{Berenbrink and Sauerwald~\protect{\cite{Berenbrink2009}}: $\bigO{\log\log(m)}$ and $\bigO{\log\log\log(m)}$ approximations for general distributions}
Building on previous results~\cite{Ross2009}, Berenbrink and Sauerwald~\cite{Berenbrink2009} consider the two approximations
\begin{align*}
  \mathcal{U}_2& := \sum_{i=1}^{m} \frac{1}{i p_i}&\text{and}& &\mathcal{U}_4& := \sum_{i=1}^{\log \log m} \frac{1}{i}\frac{1}{e^{j_i-1}p_1}\mathcal{H}_{g_{j_{i}}}
\end{align*}
where $g_i$ is the number of coupons $c$ such that $e^{i-1}<p_c/p_1 <e^i$, and $\{j_i\}_{i=1}^{\log \log m}$ is a sequence of indices such that $\{\frac{1}{e^{j_i}p_1}\mathcal{H}_{g_{j_i}}\}_{i=1}^{\log \log m}$ is decreasing. They show that $\mathcal{U}_2$ and $\mathcal{U}_4$ approximate $E[C_m]$ within $\bigO{\log\log(m)}$ and $\bigO{\log\log\log(m)}$ ratios respectively. More precisely, they show that
\begin{align*}
	&\frac{\mathcal{U}_2}{3e \log\log m}\leq E[C_m]\leq 2\cdot\mathcal{U}_2& & \text{and}  &\frac{\mathcal{U}_4}{\log\log\log m} \leq E[C_m]\leq 35\cdot \mathcal{U}_4.&
\end{align*}
Furthermore, $\mathcal{U}_2$ can be computed in polynomial time (on $n$), since there exists at most $n^{|\Sigma|}$ compositions/weights. However, the exponential growth of $m$ on $n$ limits the final precision of the approximation ratio to $\bigO{\log n}$. Finally, an efficient evaluation of $\mathcal{U}_4$ would yield a $\bigO{\log\log\log m}$ approximation in time $\Theta(\log n)$.
Unfortunately, figuring out a suitable sequence $\{j_i\}_{i=1}^{\log \log m}$ remains challenging, and seems to require knowledge over the multiplicity of coupons comparable to the one required for the application of Theorem~\ref{mainThm}.

\subsection*{Boneh and Papanicolaou~\protect{\cite{Boneh1996}}: Asymptotic estimates for truncated sequences of weighted coupons}
The authors derive general results for the asymptotics of the coupon collector problem under fairly general distributions of coupons.
They consider a fixed sequence of strictly positive weights $\balpha=\{a_k\}_{k=1}^{\infty}$, and study the truncation $\balpha_m$ of $\balpha$ to its first $m$ terms.

Their first result requires the existence of a $\xi\in]0,1]$ such that $S:=\sum_{k=1}^{\infty}\xi^{a_k}<\infty$. However, under the hypotheses of our main Theorem~\ref{mainThm}, there always exists a weight of unbounded multiplicity as $m$ goes to the infinity, and $S$ therefore diverges for any value of $\xi$.

Their second result is based on the assumption of a decreasing sequence $\balpha$.
However, many weighted distributions that satisfy hypothesis \Hypo{1} to \Hypo{3} cannot be defined by truncating a fixed decreasing sequence. For instance, suppose that for all $m$, the accessible weights are $\{\frac{2k-1}{k}\}_{k=1}^{m}\cup\{2\}$, each appearing with multiplicity $m$. It is easily checked that such a set of weights cannot be ordered into a decreasing sequence whose truncations include the families of coupons weights.

Conversely, distributions with low multiplicity satisfying their conditions are not covered by our Theorem~\ref{mainThm}. Therefore, their results and ours are complementary, and seldom overlapping.

\subsection*{Neal~\protect{\cite{Neal2008}}: The limiting distribution}
Neal studied the distribution of the waiting time. Although the results described in the article can in principle be used to assess the expectation of the waiting time,
checking the prerequisites of its main theorem turns out to be considerably more involved than checking those of Theorem~\ref{mainThm}. In particular, one has to figure out suitable sequences, 
respectively related to the expectation and variance of the distribution, from which the limiting distribution follows. This result is therefore mostly suitable to prove a conjectured distribution from a limited list of its moments. Conversely, knowledge of the expectation, as obtained from our contribution, can help figuring out suitable sequences to apply their results to.

\section{Applications to languages: the word collector}
\label{sec:application}
	\subsection{Weighted Languages}
	Let us remind some definitions introduced by Denise~\emph{et al}~\cite{Denise2010}. Let $\mathcal{L}$ be a language defined on an alphabet $\Sigma$, and let $\mathcal{L}_n$ be its restriction to words of size $n$. A positive weight $\pi_t$ is assigned to each letter $t$ of $\Sigma$. One extends these weights multiplicatively on any word $\omega\in\mathcal{L}$ such that the weight of a word $\omega$ is
	\begin{equation*}
		\pi(\omega)=\prod\limits_{t\in\omega}\pi_t.
	\end{equation*}
	This naturally defines a weighted probability distribution on $\mathcal{L}_n$, given by
	\begin{equation*}	
		\mathbb{P}[\omega] = \frac{\pi(\omega)}{\sum\limits_{\omega'\in\mathcal{L}_n}}.
	\end{equation*}
	With these definitions, $\mathcal{L}_n$ is an example of a coupon collection where each coupon is a word of $\mathcal{L}_n$. The number $m$ of coupons is the number of words of $\mathcal{L}_n$. As $m$ is now function of a $n$, all the characteristics of the weight distribution, such as $\textbf{W}_m$, will be indexed by $n$ instead of $m$.\\
	

\subsection{Verifying preconditions \Hypo{1}, \Hypo{3} and \Hypo{3} in the context of weighted languages}

Let us outline a systematic method to verify the preconditions \Hypo{1}, \Hypo{2} and \Hypo{3} for a language $\mathcal{L}$ defined over an alphabet $\Sigma=(a_1,\ldots,a_k)$. The idea is, firstly, to classify the words of the language according to their weights and find the number of words having a given weight (Step 1). Then one has to find an ordering of the different weights (Step 2). If the order cannot be found explicitly, one has to find a sufficient approximation of it (Step 3). Once this is done, the hypotheses of Theorem~\ref{mainThm} are usually easily verified.
\begin{itemize}
\item {{\bf Step 1:} \emph{Characterize the set of distinct weights.}}

The weight of a word is directly related with its composition (or sub-composition).
\begin{defn}[Compositions and sub-compositions]
	The composition of a word is the vector of occurrences of each letter within the word. More precisely, if a word $\omega$ has $x_1$ occurrence of the letter $a_1$, \ldots, $x_k$ times the letter $a_k$, irrespectively of their order, then its composition is $(x_1,\ldots,x_k)$.\\
	Suppose that $1=\pi_{a_1}=\cdots=\pi_{a_l}$ for some $l$, then the sub-composition of a word of composition $(x_1,\ldots,x_k)$ is the vector $(x_{l+1},\ldots,x_k)$, in a $(k-l)$-dimensional space, sometimes denoted $\textbf{x}$.
\end{defn}

Let us denote by $M(\textbf{x})$ the number of words of $\mathcal{L}_n$ having a given sub-composition $\textbf{x}$. By definition, any words having the same sub-composition share the same weight.
The reverse is not true in general, and words having different sub-compositions can have the same weight.
\begin{nota}
 $\Gamma_n\subset\mathbb{N}^{k-l}$ is the set of all distinct sub-compositions appearing in $\mathcal{L}_n$.
\end{nota}

\item {{\bf Step 2 :} \emph{Find a suitable ordering of weights.}}

Firstly, let us define an ordering function over $\mathcal{L}_n$, which will greatly help us characterize $\textbf{W}_n$.

\begin{defn}[Ordering function]
	Let $\phi_n$ be the application that assigns, to each sub-composition of $\Gamma_n$, the position of its weight in $\textbf{W}_n$. One has
	\begin{equation}
		\phi_n : \left\{ \begin{array}{ccl}
			\Gamma_n & \rightarrow & |\textbf{W}_n|\\
			\textbf{x} & \mapsto & i \text{, if $\pi(\textbf{x})=W_{n,i}$}.
			\end{array} \right.
	\end{equation}
	In general, this function is not bijective, therefore let us define the \textit{generalized inversed ordering function} $\tilde{\boldsymbol{\phi}}_n$ as follows :
	\begin{equation}
		\tilde{\boldsymbol{\phi}}_n : \left\{ \begin{array}{ccl}
			|\textbf{W}_n| & \rightarrow & \Gamma_n\\
			i & \mapsto & \textbf{x} \text{, if } W_{n,i}=\pi(\textbf{x})\text{ and }|\textbf{x}|=\min(|(\textbf{x}')|, W_{n,i}=\pi(\textbf{x}')),
			\end{array} \right.
	\end{equation}
where $|\textbf{x}|=x_{l+1}+\cdots+x_k$ if $\textbf{x}$ is the sub-composition $(x_{l+1},\ldots,x_k)$.
\end{defn}

With these definitions, $W_{n,i}$ and $M_{n,i}$ can be written in terms of $\phi_n$ and $\tilde{\boldsymbol{\phi}}_n$ as
\begin{align}
	W_{n,i} &= \pi(\tilde{\boldsymbol{\phi}}_n(i))&\text{and} && M_{n,i} &= \sum_{\substack{\textbf{x}\in\mathcal{L}_n,\\ \phi_n(\textbf{x})=i}} \sum\limits_{x_1+\cdots+x_l=n-|\textbf{x}|}M(\textbf{x}).\label{eqMi}
\end{align}
Sub-compositions are vectors in a $(k-l)$-dimensional space. It is easily checked that the weight of any sub-composition, found underneath the $(k-l-1)$-plane $H(\textbf{x})$ of equation $\sum_{j=l+1}^{k}x_j\log\pi_{a_i}=0$, is smaller than $\pi(\textbf{x})$, and that any sub-composition above has larger weight.
\begin{defn}
 Let $\Hyp_n(\textbf{x})\subset\Gamma_n$ be the set of sub-compositions below $H(\textbf{x})$ (all the sub-compositions that belong to $H(\textbf{x})$ have the same weight), and $S_n(\textbf{x})$ be the number of sub-compositions that belong to $H(\textbf{x})$.
\end{defn}

Then one has the following expression for $\phi_n$ :
\begin{equation}
	 \phi_n(\textbf{x})=\sum\limits_{\textbf{x'}\in\Hyp_n(\textbf{x})}\frac{1}{S_n(\textbf{x}')}.
\end{equation}
Indeed, $\phi_n$ counts the number of sub-compositions, with distinct weights, under $H(\textbf{x})$. If each weight matches a unique sub-composition, then $S_n(\textbf{x})=1$ for all $\textbf{x}$, and $\phi_n(\textbf{x})=|\Hyp_n(\textbf{x})|$.\\

\item{\bf Step 3 :} \emph{Approximate the ordering functions $\phi_n$ and $\tilde{\boldsymbol{\phi}}_n$.}

Condition \Hypo{3} directly follows from steps 1 and 2. However, conditions \Hypo{1} and \Hypo{2} require good approximations of $|\Hyp_n|$ and $S_n$. Such approximations strongly depend on the language $\mathcal{L}$ of interest, therefore we present several examples to illustrate the method.
\Comment{ ???}

\end{itemize}

\subsection{Application to specific languages}
In this part, we shall denote by $D_n$ the collection of all words of length~$n$, and assume that pairs of non-unit weights are incommensurable, which implies that sub-compositions can be bijectively associated with weights.

\Comment{A reprendre !!!!}

\subsubsection{The unconstrained language $\Sigma^*$}
Let us consider the language $\mathcal{L}=\Sigma^*$, where $\Sigma=(a_1,\ldots,a_k)$. It is worth noticing that the weighted distribution is stable upon multiplying each weight by a constant factor, therefore we assume without loss of generality that $1=\pi_{a_1}=\cdots=\pi_{a_l}$ for some $l\ge 1$, and $1<\pi_{a_{l+1}}\leq\cdots\leq\pi_{a_k}$.

Under these assumptions, one has $\Gamma_n=\{(\textbf{x}', |\textbf{x}'|\leq n)\}$. The function $\phi_n(\textbf{x})$ counts the number of sub-compositions under $H(\textbf{x})$ which belong to $\Gamma_n$. Notice that, for sufficiently large values of $n$, any sub-composition $\textbf{x}'$ belongs to $\Gamma_n$. It follows that there exists a function $\phi$ such that, for all sub-composition $\textbf{x}$ and for $n$ sufficiently large, one has $\phi_n(\textbf{x})=\phi(\textbf{x})$. From Equation~\eqref{eqMi}, one has $W_{n,i}=\pi_{a_{l+1}}^{\tilde{\boldsymbol{\phi}}_{n,1}(i)}\cdots\pi_{a_k}^{\tilde{\boldsymbol{\phi}}_{n,k-l}(i)}$, and it follows that, for sufficiently large values of $n$, one has $W_{n,i}=\pi_{a_{l+1}}^{\tilde{\boldsymbol{\phi}}_{1}(i)}\cdots\pi_{a_k}^{\tilde{\boldsymbol{\phi}}_{k-l}(i)}$.
Consequently, Condition~\Hypo{2} is verified with
\begin{align*}
	 \nu(i)&=\pi_{a_{l+1}}^{\tilde{\boldsymbol{\phi}}_{1}(i)}\cdots\pi_{a_{k}}^{\tilde{\boldsymbol{\phi}}_{k-l}(i)} &&\text{and}& \omega(n)&=1.
\end{align*}

In $\mathcal{L}_n$, the number of words of composition $(x_1,\cdots,x_k)$ is $M(x_1,\cdots,x_k)={n\choose x_1,\cdots,x_k}$, thus the number of words of sub-composition $(x_{l+1},\cdots,x_k)$ is $M(x_{l+1},\cdots,x_k)=l^{n-x_{l+1}-\cdots-x_k}{n\choose x_{l+1},\ldots,x_k}$.
Since there exists only one sub-composition $\textbf{x}$ such that $\phi(\textbf{x})=i$, then it follows from Equation~(\ref{eqMi}) that $M_{n,i}=l^{n-|\tilde{\boldsymbol{\phi}}_n(i)|}{n\choose \tilde{\boldsymbol{\phi}}_n(i)}$, where $n\choose\textbf{a}$ is the multinomial coefficient $n\choose a_1,\ldots,a_k$. Since $\phi_n=\phi$ for sufficiently large values of $n$, one has
\begin{equation}\label{eq:mni}
 M_{n,i}\underset{n\rightarrow\infty}{\sim}l^{n-|\tilde{\boldsymbol{\phi}}(i)|}{n\choose \tilde{\boldsymbol{\phi}}(i)}\underset{n\rightarrow\infty}{\sim}
 \frac{l^{n-|\tilde{\boldsymbol{\phi}}(i)|}n^{|\tilde{\boldsymbol{\phi}}(i)|}}{|\tilde{\boldsymbol{\phi}}(i)|!}.
\end{equation}

Let us now give some properties of the functions $\phi_n$ and $\phi$.
\begin{lem}\label{lemPhi}Let $S:=\sum_{j=l+1}^k\log\pi_{a_j}$, $P:=\prod_{j=l+1}^k\log\pi_{a_j}$, and let introduce a notation
$$|\textbf{x}|_{\pi}=x_{l+1}\log\pi_{a_{l+1}}+\cdots+x_k\log\pi_{a_k}.$$ Then the
following inequalities hold:
\begin{itemize}
\item[i)]  For any sub-composition $\textbf{x}$,
\begin{equation}\label{eqi}
	 \frac{|\textbf{x}|_{\pi}^{k-l}}{(k-l)!P}\leq\phi(\textbf{x})\leq\frac{(|\textbf{x}|_{\pi}+S)^{k-l}}{(k-l)!P}.
\end{equation}
\item[ii)] For all $i>0$, one has
\begin{eqnarray}\label{eqii}
\sqrt[k-l]{i(k-l)!P}-S\leq &|\tilde{\boldsymbol{\phi}}(i)|_{\pi}&\leq \sqrt[k-l]{i(k-l)!P}\\
\sqrt[k-l]{i(k-l)!\frac{P}{(\log\pi_{a_k})^{k-l}}}-\frac{S}{\log\pi_{a_k}}\leq &|\tilde{\boldsymbol{\phi}}(i)|&\leq \sqrt[k-l]{i(k-l)!\frac{P}{( \log\pi_{a_{l+1}} )^{k-l} }}.\nonumber
\end{eqnarray}
\item[iii)] For all $\textbf{x}$ and $n>0$, one has
\begin{equation}\label{eqiii}
	\phi_n(\textbf{x})\leq \phi(\textbf{x}).
\end{equation}
\item[iv)] For all $n>0$ and $i\geq1$, one has
\begin{equation}\label{eqiv}
 	\frac{\log\pi_{a_{l+1}}}{\log\pi_{a_k}}|\tilde{\boldsymbol{\phi}}(i)| \leq |\tilde{\boldsymbol{\phi}}_n(i)| \leq |\tilde{\boldsymbol{\phi}}(i)|.
\end{equation}
\end{itemize}	
\end{lem}
\begin{myproof}
\begin{itemize}
\item[i)] Remind that $\phi(\textbf{x})$ counts the number of points which are under the $(k-l-1)$-plane $H(\textbf{x})$. Equation (\ref{eqi}) just consists in bounding $\phi$ by the volume of the $(k-l-1)$-pyramid under $H(x_{l+1},\ldots,x_k)$ and the $(k-l-1)$-pyramid under $H(x_{l+1}+1,\ldots,x_k+1)$.
\item[ii)] The first equation is obtained from equation (\ref{eqi}), taking $\textbf{x}=\tilde{\boldsymbol{\phi}}(i)$. For the second equation, one uses the fact that $|\textbf{x}|\cdot \log\pi_{a_{l+1}}\leq|\textbf{x}|_{\pi}\leq|\textbf{x}|\cdot \log\pi_{a_k}$.
\item[iii)] The function $\phi_n(\textbf{x})$ counts the number of sub-compositions which are both under $H(\textbf{x})$ and belong to $\Gamma_n$, whereas $\phi(\textbf{x})$ counts the number of sub-compositions which are under $H(\textbf{x})$.
\item[iv)] For a given length $n>0$, any sub-composition is found below the $|\textbf{x}|=n$ hyperplane and, in particular, one has $|\boldsymbol{\tilde{\phi}}_n(i)|\leq n$. For some sufficiently large value of $n'>n$ , the sub-composition of $i$-th weight becomes fixed and is necessarily a sub-composition of $\Gamma_{n'}$ that did not belong to $\Gamma_n$. Consequently, this sub-composition is above the $|\textbf{x}|=n$ hyperplane, one has $|\boldsymbol{\tilde{\phi}}(i)|\geq n$ and one finally gets $|\boldsymbol{\tilde{\phi}}_n(i)|\leq|\boldsymbol{\tilde{\phi}}(i)|,\forall n>0, \forall i\ge 1$.

On the other hand, the sub-composition $\boldsymbol{\tilde{\phi}}(i)$ must be below the hyperplane $|\textbf{x}|=|\boldsymbol{\tilde{\phi}}_n(i)|_{\pi}$ otherwise its weight would be larger than the one of $\boldsymbol{\tilde{\phi}}_n(i)$. This gives $|\boldsymbol{\tilde{\phi}}_n(i)|_{\pi}\geq|\boldsymbol{\tilde{\phi}}(i)|_{\pi}$. Since any sub-composition obeys $\frac{|\textbf{x}|_{\pi}}{\log\pi_{a_{l+1}}}\geq|\textbf{x}|\geq\frac{|\textbf{x}|_{\pi}}{\log\pi_{a_k}}$, one has
\begin{equation*}
	|\boldsymbol{\tilde{\phi}}_n(i)|\geq \frac{|\boldsymbol{\tilde{\phi}}_n(i)|_{\pi}}{\log\pi_{a_k}} \geq \frac{|\boldsymbol{\tilde{\phi}}(i)|_{\pi}}{\log\pi_{a_k}} \geq \frac{\log\pi_{a_{l+1}}}{\log\pi_{a_k}}|\boldsymbol{\tilde{\phi}}(i)|,
\end{equation*}
which concludes the proof.
\end{itemize}
\end{myproof}

Combining Equations~\eqref{eq:mni} and~\eqref{eqiii}, one obtains bounds for the leading term of $M_{n,i}$, for all $i$ and as $n\to \infty$, such that
\begin{equation*}
 l^{n-|\tilde{\boldsymbol{\phi}}_n(i)|}{n\choose \tilde{\boldsymbol{\phi}}_n(i)}\leq l^{n-|\tilde{\boldsymbol{\phi}}_n(i)|}\frac{n^{|\tilde{\boldsymbol{\phi}}_n(i)|}}{|\tilde{\boldsymbol{\phi}}_n(i)|!}\leq l^{n-\frac{\log\pi_{a_{l+1}}}{\log\pi_{a_k}}|\tilde{\boldsymbol{\phi}}(i)|}\frac{n^{|\tilde{\boldsymbol{\phi}}(i)|}}{\left(\frac{\log\pi_{a_{l+1}}}{\log\pi_{a_k}}|\tilde{\boldsymbol{\phi}}(i)|\right)!}.
\end{equation*}
The convergence of $\sum\limits_{i}1/\left(\frac{\log\pi_{a_{l+1}}}{\log\pi_{a_k}}|\tilde{\boldsymbol{\phi}}(i)|\right)!$ follows from Equation~\eqref{eqii}. Therefore, \Hypo{1} is satisfied for the following choice of functions
\begin{center}
\begin{tabular}{|c|c|c|c|c|c|c|}
	\hline
				 & $F(i):=f_1(i)$ & $f_2(i)$ & $G(i):=g_1(n)$ & $g_2(n)$ & $h(i)$ & $H(i)$ \\
	\hline
	 $l=1$ & $|\tilde{\boldsymbol{\phi}}(i)|$ & & $\log n$ & & $|\tilde{\boldsymbol{\phi}}(i)|!$ & $\left(z|\tilde{\boldsymbol{\phi}}(i)|\right)!$ \\
	
	\hline
	 $l>1$ & $\log l$ & $|\tilde{\boldsymbol{\phi}}(i)|$ & $n$ & $\log n$ & $l^{|\tilde{\boldsymbol{\phi}}(i)|}|\tilde{\boldsymbol{\phi}}(i)|!$ & $l^{\left(z|\tilde{\boldsymbol{\phi}}(i)|\right)}\left(z|\tilde{\boldsymbol{\phi}}(i)|\right)!$\\
	\hline
\end{tabular}
\end{center}
where $z=\log\pi_{a_{l+1}}/\log\pi_{a_k}$. Furthermore it can be verified that \Hypo{3} is satisfied, since Equation~\eqref{eqii} gives a lower bound for $\nu(i)/F(i)$. Consequently, Theorem \ref{mainThm} applies to the weighted distribution on $\Sigma^*$, and we get.

\begin{prop}
The expected waiting time $ E[D_n]$ for obtaining all words of length $n$ in $\mathcal{L}=\Sigma^*$ admits the following asymptotic behavior:
\begin{equation*}
	E[D_n]\sim \left\{\begin{array}{cl}
	\kappa_1\cdot \mu(n)\cdot \log n &\text{if } l=1,\\
	\kappa_2\cdot  \mu(n)\cdot  n &\text{otherwise},
	\end{array}\right.
\end{equation*}
where $l$ is the number of letters of lowest weight, $\mu(n)=
\left(l+\sum_{j=l+1}^k\pi_{a_{j}}\right)^n$ is the total weight,
$\kappa_1 =  t^*\left( |\tilde{\boldsymbol{\phi}}(i)|,\lambda \right)$, and $\kappa_2 = t^*\left(\log l,\lambda \right)$ with
$\lambda = \pi_{a_{l+1}}^{\tilde{\boldsymbol{\phi}}_{1}(i)}\cdots \pi_{a_{k}}^{\tilde{\boldsymbol{\phi}}_{k-l}(i)}$.
\end{prop}

\begin{cor}
Define $p = \log( \pi_{a_1} + \cdots + \pi_{a_k} ) / \log k$, noting that $p \geq 1$ and $p = 1$ only in the uniform case.
The expected waiting time $E[C_m]$ for obtaining the $m=k^n$ words of length $n$ in ${\mathcal L} = \Sigma^*$ is asymptotically equivalent to
\begin{itemize}
\item
$\kappa_1\cdot m^p\cdot \log \log m $, if there is a single lettre of smallest weight;
\item
$ (\kappa_2/\log k)\cdot m^p \cdot \log m$, if there are at least two letters of smallest weight.
\end{itemize}
\end{cor}

\subsubsection{Motzkin words}
Motzkin words are well-parenthesized expressions featuring any number of dot characters $\bullet$. This language, denoted by $\mathcal{L}^{(m)}$, is generated by the context-free grammar
\begin{center}
	$S\rightarrow (\,S\,)\,S\;|\;\bullet\,S\;|\;\varepsilon$.
\end{center}
Here we study the expected waiting time to generate all Motzkin words of even length $n$. For the sake of readability, we replace the characters $($, $)$ and $\bullet$ by letters $a$, $\bar{a}$ and $b$ respectively. Since parentheses come in pairs, any word has equal number of occurrences of $a$ and $\bar{a}$, and the parity of the number of occurrences of $b$ is the parity of the word length. Consequently, accessible compositions for words of length $n$ are triplets $(x_a, x_{\bar{a}}, x_b)$ of the form  $(k,k,n-2k)$, with $0\leq k\leq n/2$. The number of words of size $n$ is then given by
\begin{equation*}
	M(k,k,n-2k)=\frac{1}{k+1}{2k\choose k}{n\choose 2k}.
\end{equation*}
The expected waiting time shows two types of behavior depending on whether $a$ or $\bar{a}$ have the smallest weight. To give a flavor of our result and illustrate its proof strategy, we explicitly derive two exemplary results for the cases where $1=\pi_b<\pi_a<\pi_{\bar{a}}$ and $1=\pi_a=\pi_{\bar{a}}<\pi_b$, and give the general form of the asymptotic equivalent for the weighted Coupon Collector.

\subsubsubsection{First case: $(1=\pi_b<\pi_a<\pi_{\bar{a}})$.}
Here, the sub-compositions $(x_a, x_{\bar{a}})$ are of the form $(k,k)$, $0\leq k\leq n/2$, and the associated weights are of the form $\pi_a^k\pi_{\bar{a}}^k$, increasing with $k$. Therefore one has $W_{n,i}=\pi_a^{i-1}\pi_{\bar{a}}^{i-1}$, and \Hypo{2} is satisfied with $\nu(i)=\pi_a^{i-1}\pi_{\bar{a}}^{i-1}$ and $\omega(n)=1$.
The number of words having weight $W_{n,i}$, or equivalently of sub-composition $(i-1,i-1)$, is given by
\begin{equation*}
	M_{n,i}=\frac{1}{i}{2i-2\choose i-1}{n\choose 2i-2}\underset{n\rightarrow\infty}{\sim}\frac{n^{2i-2}}{i(2i-2)!}{2i-2\choose i-1}=\frac{n^{2i-2}}{i(i-1)!^2}.
\end{equation*}
Moreover, for all $i\leq n/2$, one has $M_{n,i}\leq\frac{n^{2i-2}}{i(2i-2)!}{2i-2\choose i-1}$, and \Hypo{1} is satisfied with
\begin{center}
\begin{tabular}{|c|c|c|c|}
	\hline
		$F(i):=f_1(i)$ & $G(n):=g_1(n)$ & $h(i)$ & $H(i)$ \\
	\hline
		$2i-2$ & $\log n$ & $\frac{1}{i(i-1)!^2}$ & $\frac{1}{i(i-1)!^2}$ \\
	\hline
\end{tabular}
\end{center}
coupled with the observation that $\sum_i 1/i(i-1)!^2$ converges.
The verification of \Hypo{3} is immediate, and applying Theorem~\ref{mainThm} readily gives the following result.

\begin{prop}
The expected waiting time of the full collection of weighted Motzkin words of even length $n$, under the configuration $1=\pi_b<\pi_a<\pi_{\bar{a}}$, admits the following asymptotic behavior:
\begin{equation*}
	E[D_n]\sim \kappa\cdot\mu(n)\cdot\log n
\end{equation*}
where $\kappa=t^*\left(2i-2,\pi_a^{i-1}\pi_{\bar{a}}^{i-1}\right)$ and $\mu(n)=\sum_{k=0}^{n/2}\frac{1}{k+1}{2k\choose k}{n\choose 2k}\pi_a^{k}\pi_{\bar{a}}^k$.
\end{prop}

\subsubsubsection{Second case: $(1=\pi_a=\pi_{\bar{a}}<\pi_b)$.} In this second case, the sub-compositions $(x_b)$ are of the form $(n-2k)$, for $0\leq k\leq n/2$, and the weight of a word increases with the number of occurrences of $b$. Consequently, one has $W_{n,i}=\pi_b^{2(i-1)}$, and \Hypo{2} is satisfied with
$\nu(i)=\pi_b^{2(i-1)}$ and $\omega(n)=1$.
Furthermore, if $(n-2k)$ is the sub-composition of the $i$-th weight, then $n-2k=2(i-1)$, leading to $k=n/2-(i-1)$ and one finally has
\begin{equation*}
	M_{n,i}=\frac{1}{\frac{n}{2}-(i-1)+1}{n-2(i-1)\choose \frac{n}{2}-(i-1)}{n\choose n-2(i-1)}\underset{n\rightarrow\infty}{\sim}2^n\frac{n^{2(i-1)-\frac{3}{2}}}{\sqrt{\pi}2^{2(i-1)-\frac{3}{2}}(2(i-1))!}.
\end{equation*}
Finally, one has $M_{n,i}\leq2^n\frac{n^{2(i-1)-\frac{3}{2}}}{\sqrt{\pi}2^{2(i-1)-\frac{3}{2}}(2(i-1))!}$, for $i\leq n/2$, and \Hypo{1} is satisfied with
\begin{center}
\begin{tabular}{|c|c|c|c|c|c|}
\hline
		$f_1(i)$ & $f_2(i)$ & $g_1(n)$ & $g_2(n)$ & $h(i)$ & $H(i)$ \\
	\hline
		$\log 2$ & $2(i-1)-\frac{3}{2}$ & $n$ & $\log n$ & $\sqrt{\pi}2^{2(i-1)-\frac{3}{2}}(2(i-1))!$ & $\sqrt{\pi}2^{2(i-1)-\frac{3}{2}}(2(i-1))!$ \\
	\hline
\end{tabular}
\end{center}
 since $\sum\limits_i 1/2^{2i}(2(i-1))!$ converges.
Again, verifying \Hypo{3} is immediate.
\begin{prop}
The expected waiting time of the full collection of weighted Motzkin words of even length $n$, under the configuration $1=\pi_a=\pi_{\bar{a}}<\pi_b$, admits the following asymptotic behavior:
\begin{equation*}
	E[D_n]\sim \kappa\cdot \mu(n)\cdot n
\end{equation*}
where $\kappa=t^*\left(\log 2,\pi_b^{2(i-1)}\right)$ and $\mu(n)=\sum\limits_{k=0}^{n/2}\frac{1}{k+1}{2k\choose k}{n\choose 2k}\pi_b^{n-2k}$.
\end{prop}

This approach can be extended to any relative positioning of $\pi_a$, $\pi_{\bar{a}}$ and $\pi_b$. The symmetrical roles played by the letters $a$ and $\bar{a}$, allow for a restriction, without loss of generality, to cases where $\pi_{\bar{a}} \leq \pi_a$. Also, singularity analysis can be applied to the generating function of the weighted Motzkin language, giving $\mu_n \sim \kappa \cdot \rho^{-n} \cdot n^{-3/2}$, with $\rho=(\pi_b+2\sqrt{\pi_a}{\pi_{\bar{a}}})^{-1}$.
\begin{prop}
The expected waiting time for generating all Motzkin words of length $n$ obeys:
\[
E[D_n] \sim \left\{
\begin{array}{cll}
  \kappa \frac{\rho^{-n}}{\sqrt{n}} &\text{with }\rho = \frac{\sqrt{\pi_a}}{\pi_b + 2 \sqrt{\pi_a \pi_{\bar{a}}}}& \text{if }\; \pi_b^2 > \pi_a \pi_{\bar{a}}, \\
  \kappa' \frac{\rho'^{-n}}{n\sqrt{n}} \log n &\text{with }\rho' = \frac{\pi_b} {\pi_b + 2 \sqrt{ \pi_a \pi_{\bar{a}}  } }& \text{if }\; \pi_b^2 < \pi_a \pi_{\bar{a}},
\end{array}
\right.
\]
 where $\kappa$ and $\kappa'$ are constants of $n$ which can be explicitly computed (and depends on the relative positions of the weights).
\end{prop}

\begin{cor}
Let $m$ be the number of Motzkin words of length $n$ ($m\sim 3 (\sqrt{3}/2 \sqrt{\pi}) 3^n n^{-3/2}$).
The expected waiting time for generating the complete collection of $m$ words obeys
\[
E[C_n] \sim \left\{
\begin{array}{cll}
   \kappa \cdot m^p  \cdot \log(m)^{\frac{3p-1}{2}} &\text{with }p= \frac{\log ( \pi_b + 2 \sqrt{\pi_a \pi_{\bar{a}}  } ) - \log \sqrt{\pi_a}}{\log3} & \text{if }\; \pi_b^2 > \pi_a \pi_{\bar{a}}, \\
  \kappa'  \cdot m^{p'}  \cdot \log(m)^{\frac{3(p'-1)}{2}} \cdot \log \log m &\text{with }p' = \frac{\log (\pi_b + 2 \sqrt{ \pi_a \pi_{\bar{a}}  }) - \log \pi_b}{\log 3}& \text{if }\; \pi_b^2 < \pi_a \pi_{\bar{a}},
\end{array}
\right.
\]
for constants $\kappa$ and $\kappa'$ that can be explicitly computed (and depend on the relative positions of the weights).
\end{cor}

\subsubsection{RNA secondary structures}
\usetikzlibrary{positioning}
\begin{figure}\label{rna}
  \scalebox{.8}{\input{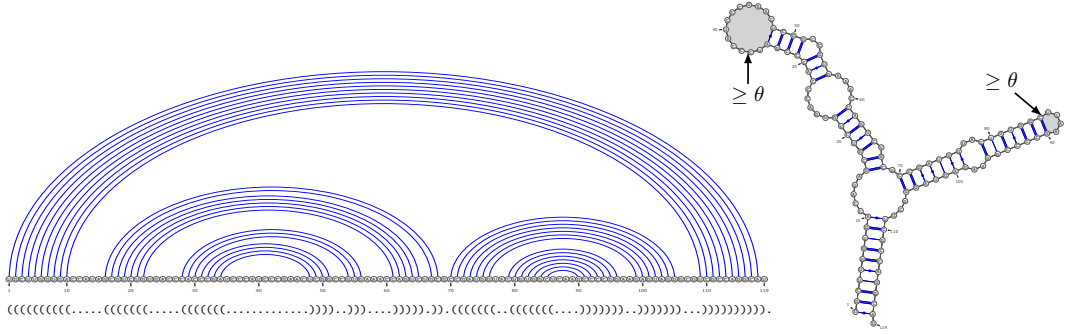}}
  \caption{Secondary structure of a 5s ribosomal RNA. A well-parenthesized expression (lower-left) unambiguously defines a set of matching position (upper-left) which \emph{folds} into a projection of a three-dimensional conformation of the molecule. The latter representation illustrates the relationship between the $\ge\theta$ steric constraint and the absence of sharp turns.\label{fig:arn}}
\end{figure}

Through an adaptation of Viennot \emph{et al}~\cite{Chaumont1983}, secondary structures can be generated by a grammar:
\begin{align*}
	S&\rightarrow (\,S_{\geq \theta}\,)\,S\;|\;\bullet\,S\;|\;\varepsilon& \text{and}&&
 S_{\geq\theta}&\rightarrow (\,S_{\geq\theta}\,)\,S\;|\;\bullet\,S_{\geq\theta}\;|\;\bullet^\theta,
\end{align*}
where $\theta$ is the minimal distance between matching parenthesis, enforcing steric constraints.
The connection between the secondary structure and the conformations of an RNA sequence is illustrated by Figure~\ref{fig:arn}:
Matching parentheses represent base-pairs, or interacting pairs of nucleotides 
mediated by hydrogen bonds.
Such base-pairs are known to stabilize a secondary structure, thus decreasing its free-energy. In this model, we consider
a simple free-energy model proposed by Nussinov\cite{nussinov80} which assigns a $-1$ kcal/mol contribution to each base-pair.
The free-energy $E(S)$ of a secondary structure $S$ is then inherited additively by summing the individual contributions of its base-pairs.

One can assume a Boltzmann distribution on the set of secondary structures, where the probability of any secondary structure $S$ of length $n$ is
proportional to its Boltzmann factor $e^{-E(S)/RT}$, with $R$ the gas constant and $T$ the temperature.
Such a non-deterministic perspective over the RNA folding process is fundamental to a recent paradigm shift in RNA structure prediction~\cite{DiLa03} based on random generation.
In the worst-case scenario, the complexity of this algorithm is equivalent to a coupon collector for Boltzmann weighted secondary structures.
It is then worth noticing that the Boltzmann distribution is just a special case of a weighted distribution, where a neutral weight $1$ is assigned to unpaired positions,
and a weight $e^{1/RT}$ to each pair of matching parentheses.

Again in this example, we replace the characters $($, $)$ and $\bullet$ by letters $a$, $\bar{a}$ and $b$ respectively.
Let us denote by $\mathcal{L}^{(rna)}$ the language of RNA secondary structure. For the sake of simplicity, let us assume, without loss of generality, that $1=\pi_b<\pi_a<\pi_{\bar{a}}$, with $\pi_a\cdot \pi_{\bar{a}}=e^{1/RT}$. The compositions are triplets $(x_a,x_{\bar a},x_b)$  of the form $(k,k,n-2k)$, $0\leq k\leq n/2$. The number of words of size $n$ having $p$ \textit{plateaux} and $k$ occurrences of $a$ is given by $1$ if $(p,k)=(0,0)$, and $s_{n,k,p,\theta}=\frac{1}{k}{k\choose p}{k\choose p-1}{n-\theta p\choose 2k}$ otherwise. Consequently, the number of words having a given composition $(k,k,n-2k)$ is such that
\begin{equation*}
	M(k,k,n-2k)=\delta_{k,0}+\sum\limits_{p=1}^{\left\lfloor \frac{n-2k}{\theta}\right\rfloor}s_{n,k,p,\theta}=\delta_{k,0}+\sum\limits_{p=1}^{\left\lfloor \frac{n-2k}{\theta}\right\rfloor}\frac{1}{k}{k\choose p}{k\choose p-1}{n-\theta p\choose 2k}
\end{equation*}
where $\delta$ is the Kronecker symbol ($\delta_{a,b}=1$ if $a=b$, and $0$ otherwise). Since $1=\pi_b<\pi_a<\pi_{\bar{a}}$, the weights of words are increasing with the number of $\bar{a}$.
 It follows that $W_{n,i}=\pi_a^{(i-1)}\pi_{\bar{a}}^{(i-1)}$, and \Hypo{2} is satisfied with $\nu(i)=\pi_a^{(i-1)}\pi_{\bar{a}}^{(i-1)}$ and $\omega(n)=1$.

Moreover, the multiplicity of the weight $W_{n,i}$ is the number of words having sub-composition $(x_a,x_{\bar a})$ of the form $(i-1,i-1)$, and is given by
\begin{equation*}
	M_{n,i}=\delta_{i-1,0}+\sum\limits_{p=1}^{\left\lfloor \frac{n-2(i-1)}{\theta}\right\rfloor}\frac{1}{(i-1)}{i-1\choose p}{i-1\choose p-1}{n-\theta p\choose 2(i-1)}\underset{n\rightarrow\infty}{\sim}\frac{n^{2(i-1)}}{i(i-1)!^2}.
\end{equation*}
Indeed, for large values of $n$, the scope of the sum above can be limited to $p\in[1,i-1]$ since any term such that $p>(i-1)$ has null contribution.

One also has $M_{n,i}\leq2\frac{n^{2(i-1)}}{i(i-1)!^2}$, for all $i$,  thus \Hypo{1} is satisfied with
\begin{center}
\begin{tabular}{|c|c|c|c|}
	\hline
		$f_1(i)$ & $g_1(n)$ & $h(i)$ & $H(i)$ \\
	\hline
		$2(i-1)$ & $\log n$ & $i(i-1)!^2$ & $\frac{i}{2}(i-1)!^2$ \\
	\hline
\end{tabular}
\end{center}
where $\sum\limits_i 1/H(i)$ obviously converges, and the verification of \Hypo{3} is immediate.
Setting
\[
\mu(n)=\sum\limits_{k=0}^{\left\lfloor \frac{n}{2}\right\rfloor}\left[\delta_{k,0}+\sum\limits_{p=1}^{\left\lfloor \frac{n-2k}{\theta}\right\rfloor}\frac{1}{k}{k\choose p}{k\choose p-1}{n-\theta p\choose 2k}\right]
(\pi_a \pi_{\bar{a}})^k,
\]
one verifies, e.g. from the strong-connectivity of the grammar~\cite{Drmota97}, that
\[
\mu_n \sim
c\cdot \left( \frac{1}{\rho^{\theta}} \right)^n n^{-3/2}
\]
where $c$ is a constant, and $\rho^{\theta}$ is the dominant singularity of $\sum_{n\ge 0} \mu(n)\cdot z^n$.
\begin{prop}
The expected waiting time for the collection of Boltzmann-factor weighted RNA secondary structures of length $n$, assuming $1=\pi_b<\pi_a<\pi_{\bar{a}}$, admits the following asymptotic behavior:
\begin{equation*}
	E[D_n]\sim \kappa \cdot \frac{\rho_\theta^{-n}}{n\sqrt{n}} \cdot\log n,\;  \forall\theta\in \mathbb{N}^+
\end{equation*}
where $\kappa=t^*\left(2(i-1),(\pi_a \pi_{\bar{a}})^{i-1} \right)$
and, setting $q = \pi_a \pi_{\bar{a}}$, $\rho_\theta$ is the smallest positive real solution of
\[
1-4z+(6-2 q )z^2+4(q-1)z^3+(1-2q)z^4-2q z^{\theta+2}+4q z^{\theta+3}-2q (1+q )z^{\theta+4}+q^2z^{2\theta+4} =0.
\]
\end{prop}

\begin{cor}
Define $\eta_\theta$ as the smallest positive solution of the equation
\[
1-4z+4z^2 - z^4 -2 z^{\theta+2} + 4 z^{\theta+3} - 4 z^{\theta+4} + z^{2\theta+4}=0.
\]
Then the number $m$ of RNA structures of length $n$ is asymptotically equal to $\lambda\cdot\eta_\theta^n \cdot n^{-3/2}$, and the asymptotic waiting time of the full collection is given by
\[
E[C_m] \sim
\kappa \cdot m^p \cdot (\log m)^{3p/2}\cdot \log \log m,
\]
where $p = -\frac{ \log \rho_\theta}{\log \eta_\theta}$, and $\lambda$ and $\kappa$ are constants that can be fully specified.
\end{cor}

\subsubsection{A non strongly-connected language}
Let us finally consider the language $\mathcal{L}^{(nc)}$ over an alphabet $\{a,\bar{a},b\}$ and generated by the grammar
\begin{align*}
	S&\rightarrow \bar{a}\,S\,b\,U\;|\;\varepsilon& \text{and}&&
 U&\rightarrow a\,U\,b\,U\;|\;\varepsilon.
\end{align*}
It is worth noticing that this grammar is not strongly connected, and the distributions of letters may therefore be untypical (non-normal and/or expectation/variance
not in $O(n)$~\cite{Drmota97}). Here, this grammar models binary trees, whose leftward edges along the leftmost branch are marked by a dedicated letter ${\bar{a}}$, and any other leftward (resp. rightward) edge is marked by $a$ (resp. $b$).

The restriction of $\mathcal{L}^{(nc)}$ to words of odd length is empty, thus we only study the word collector on even sizes.
The structure of this grammar is such that each word of size $n$ has exactly $n/2$ occurrences of the letter $b$, and compositions are therefore triplets
$(x_a,x_{\bar a},x_b)$  of the form $(n/2-k,k,n/2)$, for $1\leq k\leq n/2$. An elementary computation shows that the number of words of a given composition is
\begin{equation*}
	M(n/2-k,k,n/2)={n-k-1\choose n/2-1}-{n-k-1\choose n/2}.
\end{equation*}
The expected waiting time depends on the relative position of the weights associated with letters, leading to different behaviors.
Let us illustrate the approach on one out of the 9 possible configurations, such that $1=\pi_b<\pi_a<\pi_{\bar{a}}$.

In this case, the sub-compositions are pairs $(x_a,x_{\bar a})$  of the form $(n/2-k,k)$, for $1\leq k\leq n/2$. Moreover, since the weight of the word increases with the number of $\bar{a}$,
then one has $W_{n,i}=\pi_a^{n/2-i} \pi_{\bar{a}}^{i}$, and \Hypo{2} is therefore satisfied with $\nu(i)=\left( {\pi_{\bar{a}}} / {\pi_a} \right)^i$ and $\omega(n)=\pi_a^{n/2}$.
\begin{rem}
The influence of the configuration (ordering of the weights) only appears in the definition of the functions $\nu$ and $\omega$.
The function $\omega$ may become constant (equal to~1) when either $\pi_b=\pi_a=1$ or $\pi_b=\pi_{\bar{a}}=1$.
\end{rem}
Now the number of words having the $i$-th weight, i.e. the sub-composition $(i,n/2-i)$, is given by
\begin{equation}
	M_{n,i}={n-i-1\choose n/2-1}-{n-i-1\choose n/2}\underset{n\rightarrow\infty}{\sim}2^{n-i}n^{-\frac{3}{2}}i\sqrt{\frac{2}{\pi}}.\label{eq:nonstrongconn}
\end{equation}
Since $M_{n,i}\leq 2^{n-i+1} n^{-\frac{3}{2}} i \sqrt{2/\pi} $ for all $1\leq i\leq n/2$ and $\sum\limits_i i/2^i$ converges, the condition \Hypo{1} is satisfied for the following functions:
\begin{center}
\begin{tabular}{|c|c|c|c|c|c|}
	\hline
		$F(i):=f_1(i)$ & $f_2(i)$ & $G(n):=g_1(n)$ & $g_2(n)$ & $h(i)$ & $H(i)$ \\
	\hline
		$\log 2$ & $-\frac{3}{2}$ & $n$ & $\log n$ & $\frac{2^i}{i}\sqrt{\frac{\pi}{2}}$ & $\frac{2^{i-1}}{i}\sqrt{\frac{\pi}{2}}$ \\
	\hline
\end{tabular}
\end{center}
The verification of \Hypo{3} is immediate.
\\
From \eqref{eq:nonstrongconn}, one has $\mu(n)=\sum\limits_{k=1}^{n/2}\left[{n-k-1\choose n/2-1}-{n-k-1\choose n/2}\right]\pi_a^{\frac{n}{2}-k}\pi_{\bar{a}}^k$, whose
 asymptotic behaviour obeys
\[
\mu(n) \sim
2\sqrt{2} \frac{\pi_a \pi_{\bar{a}}}{(2 \pi_a - \pi_{\bar{a}}  )^2} \left( { 2 \sqrt{ \pi_a }}  \right)^n  n^{-3/2}.
 \]
\begin{prop}
The expected waiting time for obtaining all words in $\mathcal{L}^{(nc)}$ of even length $n$, under the configuration $1=\pi_b<\pi_a<\pi_{\bar{a}}$, admits the following asymptotic behavior:
\begin{equation*}
	E[D_n]\sim \kappa  \cdot  \frac{2^n}{\sqrt{n}},
\end{equation*}
where
$\kappa= t^*\left(\log2,\left({\pi_{\bar{a}}}/{\pi_a}\right)^i\right)\cdot
\frac{2 \sqrt{2} \pi_a \pi_{\bar{a}} }{ (2 \pi_a - \pi_{\bar{a}}  )^2 }$.
\end{prop}

\begin{cor}
Let $m$ be the number of words of even length $n$ in $\mathcal{L}^{(nc)}$, asymptotically equivalent to
$2\sqrt{2/\pi}\cdot 2^n\cdot n^{-3/2}$.
The expected waiting time of the full collection is
\[
E[C_m] \sim \kappa \cdot m \cdot( \log m )^{5/2} ,
\]
where $\kappa$ is a constant that can be explicitly computed.
\end{cor}

Again, these results can be extended to any relative ordering of $\pi_a$, $\pi_{\bar{a}}$ and $\pi_b$, and one obtains the following result.
\begin{prop}
The expected waiting time for all words of even length $n$  in $\mathcal{L}^{(nc)}$ is equivalent to
\[
E[D_n] \sim \left\{\begin{array}{cl}
  \kappa\cdot\frac{2^n}{\sqrt{n}} &  \text{if }\pi_a=1, \text{ or }1=\pi_b \leq \pi_a < \pi_{\bar{a}},\\
  \kappa'\cdot\left(\frac{\pi_a}{\pi_{\bar{a}}} \right)^{n/2}\cdot 2^n\cdot \frac{\log n }{n\sqrt{n}} & \text{otherwise,}
\end{array}\right.
\]
where $\kappa$ and $\kappa'$ are constants that can be explicitly computed.
\end{prop}

\begin{cor} Let $m$ be the number of words of even length $m$ in $\mathcal{L}^{(nc)}$.
Then the expected waiting time of the complete collection
is asymptotically equal to
\[
E[C_m] \sim \left\{
  \begin{array}{cl}
    \kappa\cdot m^2\cdot  \left( \log m \right)^{5/2} & \text{if }\pi_a=1\text{ or }\pi_b =1 \leq \pi_a < \pi_{\bar{a}},\\
    \kappa'\cdot  m^{2+q}\cdot  \left( \log m  \right)^{2+q/2}\cdot  \log \log m & \text{otherwise, with }q= \log_2 (\pi_a / \pi_{\bar{a}})\\
  \end{array}
\right.
\]
where $\kappa$ and $\kappa'$ are constants that can be explicitly computed.
\end{cor}

\section{Conclusion}
\label{sec:conclusion}

In this extended abstract, we studied a language generalization of the ubiquitous Coupon Collector Problem.
Focusing on collections of weighted coupons having large multiplicities, we contributed a new theorem that relates
the asymptotic waiting time of the full to the growth of the multiplicity of coupons of a given weight.
We compared the novelty of the contribution against pre-existing work on the subject.
We discussed the application of our theorem to weighted languages in general, and particularly on four languages showing different properties
(rational vs context-free, simple-type vs non-square-root singularities, limited vs parameterized alphabet\ldots).
\Comment{ Changer ordre ??}
\Comment{ Discuter les changements de rgime (Difficiles  observer numriquement)}

Quite interestingly, our study of four illustrative examples reveals a large variety of expressions for the waiting-time.
As a function of the word length $n$, we observed waiting times of the form $\kappa\cdot \mu(n)\cdot n$ and $\kappa\cdot \mu(n)\cdot \log(n)$, depending essentially
on the multiplicity of the smallest weights. As a function of the number of coupons $m$, we obtained estimates of the general form  $\kappa\cdot m^p \cdot (\log m)^{q}\cdot (\log \log m)^{\theta}$,
where $p$ and $q$ are irrational numbers and $\theta\in\{0,1\}$. Such a diversity partly not only arises from differences regarding the nature of the asymptotical growth within the language, but
also reflects subtle differences in the accumulation of the contributions of the least probable words. To our opinion, this illustrates the versatility of the method, and hints toward a significant amount of work being required, in the case of approximations~\cite{Berenbrink2009}.

Perhaps the main limitation of our work lies in the prerequisites of Theorem~\ref{mainThm}.
As shown in Section~\ref{sec:application}, verifying these -- technically involved -- conditions is already made easier in the context of languages.
However, one could imagine characterizing broad classes of languages that automatically verify these conditions.
For instance, conditions of aperiodicity (a.k.a. lattice-type~\cite{Flajolet2009}) and strong-connectivity of a context-free grammar are known to ensure typical asymptotic growths,
both for the total number of words, their cumulated weight and the total number of words of a given composition~\cite{Drmota97}.
We hope that such conditions, possibly in addition to other easily-checkable properties, could provide a sufficient set of conditions for a given regime.

Another natural extension may generalize the results to multi-parameterized combinatorial classes, as generated by decomposable combinatorial classes~\cite{flajoletcalculus}.
The main difficulties behind such an extension are related to the variety of asymptotic growths that may appear, e.g. for the substitution construct, in addition to an increased level of difficulty for determining the number of words of a given composition/weight. This both motivates a further relaxation of the -- sufficient but not necessary -- conditions of Theorem~\ref{mainThm}, along with a study of accessible asymptotics for the growth of coefficients in multivariate generating functions.

\Comment{ Si le temps le permet, discuter un langage non-captur par les conditions du thorme ($(a,b)^*\cap(c,d)^*$).}

\acknowledgements
The authors wish to thank an anonymous reviewer for suggesting a more intuitive presentation of our main result.
This work was supported by the French \emph{Agence Nationale de la Recherche}
through the BOOLE {\tt ANR 09 BLAN 0011} (JDB and DG) and
MAGNUM {\tt ANR 2010 BLAN 0204} (YP) grants.

\bibliographystyle{amsplain}
\bibliography{biblio}

\appendix
\section{Proof of Theorem \ref{mainThm}}
\label{sec:appendix}

For the proof of the theorem, we need the following lemma.

\begin{lem}\label{lemfg}
		Let $E\subset\mathbb{N}^*$. Let $f$ and $g$ be two non-zero positive functions on $E$, such that if $E$ is not finite, $\lim\limits_{x\rightarrow\infty}\frac{g(x)}{f(x)}=+\infty$. Then, \\
			- $\exists\ t^\ast(f,g)>0$ such that\\

			\begin{tabular}{lcr}
				(1) $\forall 0\leq t<t^\ast(f,g)$,& $\exists\ x_0\in E$,& $f(x_0)-t g(x_0)>0$\\
				(2) $\forall t>t^\ast(f,g)$,& $\forall\ x\in E$,& $f(x)-t g(x)<0$\\
			\end{tabular}\\
		
			\noindent - $\exists\ x_1\in \mathbb{N}^*$ such that\\
		
			\begin{tabular}{lcr}
				(3) $f(x_1)-g(x_1)t=\max\limits_{x\in E}(f(x)-tg(x))$
			\end{tabular}
\end{lem}

\begin{myproof}
	Throughout the proof, $f(x)-tg(x)$ is seen as a function of $x$ with a parameter $t$.\\
	Let us define $t_x=\frac{f(x)}{g(x)}$, for all $x\in E$. $\forall\ t<t_x$, $f(x)-tg(x)>0$ and $\forall t>t_x$, $f(x)-tg(x)<0$. If $E$ is finite, it is obvious that $t_x$ reaches its maximum, i.e. there is $X\in E$ such that $t_X=\max\limits_{x\in E}(t_x)$. This property is still true when $E$ is not finite because $t_x\rightarrow0$ as $x\rightarrow\infty$. Then, $(1)$ and $(2)$ are satisfied, taking $t^\ast(f,g)=t_X$.\\
	If $E$ is finite, it is obvious that $f(x)-tg(x)$ reaches its maximum for all $t>0$. If $E$ is not finite, using the fact that $\lim\limits_{x\rightarrow\infty}\frac{g(x)}{f(x)}=+\infty$, we have $\forall t>0$, $f(x)-tg(x)\underset{x\rightarrow\infty}{\rightarrow}-\infty$. Then $f(x)-tg(x)$ reaches its maximum, i.e. there is $x_1\in E$ such that $f(x_1)-(x_1)t=\max\limits_{x\in E}(f(x)-g(x)t)$, which proves $(3)$.
\end{myproof}

\noindent \textbf{Proof of the theorem.\\}
Let us suppose that $\textbf{W}_m$ satisfies $\textbf{H1}$, $\textbf{H2}$, and $\textbf{H3}$. From equation (\ref{eq:couponcollector}), we have
		\begin{equation*}
			E[C_m] = \int_{0}^{\infty}\left[1-\prod\limits_{i=1}^{\ws}\left(1-e^{-\frac{W_{m,i}}{\mu_m}u}\right)^{M_{m,i}}\right]du.
		\end{equation*}\\
	The substitution $u \frac{\omega(m)}{\mu_{m}\sum\limits_{j=1}^{p}g_j(m)}\rightarrow t$ gives
	\begin{eqnarray*}
		E[C_m] &=& \frac{\mu_{m}}{\omega(m)}\sum\limits_{j=1}^{p}g_j(m)\int_{0}^{\infty}\left[1-\prod\limits_{i=1}^{\ws}\left(1-e^{-t \frac{W_{m,i}}{\omega(m)}\sum\limits_{j=1}^{p}g_j(m)}\right)^{M_{m,i}}\right]dt\\
		&=&
\frac{\mu_{m}}{\omega(m)}\sum\limits_{j=1}^{p}g_j(m)\int_{0}^{\infty}\left[1-\exp\left(\sum\limits_{i=1}^{\ws}M_{m,i}\log\left(1-e^{-\frac{W_{m,i}}{\omega(m)}t\sum\limits_{j=1}^{p}g_j(m)}\right)\right)\right]dt.
	\end{eqnarray*}
	From \textbf{H1}, we have $\sum\limits_{j=1}^{p}g_j(m)\sim g_1(m)$. To conclude, we have to show that the integral converges when $m$ goes to infinity. First, we show that the integral from 0 to $t^\ast(f_1,c)$ converges to $t^\ast(f_1,c)$. Then, we show that the remaining integral converges to $0$.\\
	
	$\bullet$ From Lemma \ref{lemfg}, applied to $E$, and \textbf{H3} (if $\ws\underset{m\rightarrow\infty}{\rightarrow}\infty$), there is $i_0\in E$ such that $f_1(i_0)-\nu(i_0)t>0$. Moreover, from \textbf{H2}, for $m$ sufficiently large, $W_{m,i_0}=\nu(i_0)\omega(m)$. Then
	\begin{eqnarray*}
		\sum\limits_{i=1}^{\ws}M_{m,i}\log\left(1-e^{-\frac{W_{m,i}}{\omega(m)}t\sum\limits_{j=1}^{p}g_j(m)}\right)&\leq& -\sum\limits_{i=1}^{\ws}M_{m,i}e^{-\frac{W_{m,i}}{\omega(m)}t\sum\limits_{j=1}^{p}g_j(m)}\\
&\leq& -M_{m,i_0}e^{-\frac{W_{m,i_0}}{\omega(m)}t\sum\limits_{j=1}^{p}g_j(m)} = -M_{m,i_0}e^{-\nu(i_0)t\sum\limits_{j=1}^{p}g_j(m)}.
	\end{eqnarray*}
	From \textbf{H1}, for $m$ sufficiently large, $M_{m,i_0}\geq \frac{1}{2}\frac{e^{\sum\limits_{j=1}^{p}f_j(i_0)g_j(m)}}{h(i_0)}$. Then,
	\begin{equation*}
 \sum\limits_{i=1}^{\ws}M_{m,i}\log\left(1-e^{-\frac{W_{m,i}}{\omega(m)}t\sum\limits_{j=1}^{p}g_j(m)}\right) \leq -\frac{e^{\sum\limits_{j=1}^{p}\left(f_j(i_0)-\nu(i_0)t\right)g_j(m)}}{2h(i_0)}.
	\end{equation*}
	As $f_1(i_0)-\nu(i_0)t>0$ and $g_j(m)=o(g_1(m))$ for all $j>1$, $\sum\limits_{j=1}^{p}\left(f_j(i_0)-\nu(i_0)t\right)g_j(m)\underset{m\rightarrow\infty}{\rightarrow}+\infty$. Then,
	\begin{equation*}	\sum\limits_{i=1}^{\ws}M_{m,i}\log\left(1-e^{-\frac{W_{m,i}}{\omega(m)}t\sum\limits_{j=1}^{p}g_j(m)}\right)\underset{m\rightarrow\infty}{\rightarrow}-\infty,
	\end{equation*}
	and
	\begin{equation*}	\prod\limits_{i=1}^{\ws}\left(1-e^{-\frac{W_{m,i}}{\omega(m)}t\sum\limits_{j=1}^{p}g_j(m)}\right)^{M_{m,i}}\underset{m\rightarrow\infty}{\rightarrow}0.
	\end{equation*}
	This leads to
	\begin{equation}
	\label{eq1}	
	\int_{0}^{t^\ast(f_1,\nu)}\left[1-\prod\limits_{i=1}^{\ws}\left(1-e^{-\frac{W_{m,i}}{\omega(m)}t\sum\limits_{j=1}^{p}g_j(m)}\right)^{M_{m,i}}\right]dt \underset{m\rightarrow\infty}{\rightarrow}t^\ast(f_1,\nu).
	\end{equation}
	\\
	By definition, $W_{m,i}/\omega(m)$ is increasing in $i$, and from \textbf{H2}, for $m$ sufficiently large, $W_{m,1}/\omega(m)=\nu(1)$.
	Moreover, $\sum\limits_{j=1}^{p}g_j(m)\sim g_1(m)\rightarrow+\infty$, from \textbf{H1}. Then, for $m$ sufficiently large, $\forall t>t^\ast(f_1,\nu)$, $e^{-\frac{W_{m,i}}{\omega(m)}t\sum\limits_{j=1}^{p}g_j(m)}<\frac{1}{2}$. Using $\log(1-x)\geq-2x$ for all $x\leq1/2$, we have
	\begin{equation*}
		\sum\limits_{i=1}^{\ws}M_{m,i}\log\left(1-e^{-\frac{W_{m,i}}{\omega(m)}t\sum\limits_{j=1}^{p}g_j(m)}\right)\geq -2\sum\limits_{i=1}^{\ws}M_{m,i}e^{-\frac{W_{m,i}}{\omega(m)}t\sum\limits_{j=1}^{p}g_j(m)}.
	\end{equation*}
	From \textbf{H1}, we have that for all $i$, $M_{m,i} \leq \frac{e^{\sum\limits_{j=1}^{p}f_j(i)g_j(m)}}{H(i)}$. From \textbf{H2}, for all $i$, $W_{m,i}\geq \nu(i)\omega(m)$. Thus,
	\begin{equation*}
		\sum\limits_{i=1}^{\ws}M_{m,i}\log\left(1-e^{-\frac{W_{m,i}}{\omega(m)}t\sum\limits_{j=1}^{p}g_j(m)}\right)\geq -2\sum\limits_{i=1}^{\ws}\frac{e^{\sum\limits_{j=1}^{p}\left(f_j(i)-\nu(i)t\right)g_j(m)}}{H(i)}.
	\end{equation*}
	$\forall t>t^\ast(f_1,\nu)$, we have $f_1(i)-\nu(i)t<0$ for all $i\leq \ws$. From \textbf{H3}, there exists $K>0$ such that for all $1<j\leq p$, for all $i\leq\ws$ and for all $t>t^\ast(f_1,\nu)$, $\left(f_j(i)-\nu(i)t\right)\leq K$. Then, $\forall i\in E$,
	\begin{equation*}
		\sum\limits_{j=1}^{p}\left(f_j(i)-\nu(i)t\right)g_j(m) \leq K\sum\limits_{j=2}^{p}g_j(m)+(f_1(i)-\nu(i)t)g_1(m).
	\end{equation*}
  For all $j\neq 1$ we have $g_j=o(g_1)$. Thus, for $m$ sufficiently large, $\sum\limits_{j=1}^{p}\left(f_j(i)-\nu(i)t\right)g_j(m) \leq 2(f_1(i)-\nu(i)t)g_1(m)$.
Then,
	\begin{eqnarray*}
		\sum\limits_{i=1}^{\ws}M_{m,i}\log\left(1-e^{-\frac{W_{m,i}}{\omega}t\sum\limits_{j=1}^{p}g_j(m)}\right) &\geq&
		-2\sum\limits_{i=1}^{\ws}\frac{e^{2(f_1(i)-\nu(i)t)g_1(m)}}{H(i)}\\
		&\geq& -2e^{2g_1(m)\max\limits_{i\in E}(f_1(i)-\nu(i)t)}\sum\limits_{i=1}^{\ws}\frac{1}{H(i)}.
	\end{eqnarray*}
	From \textbf{H1}, there is $C>0$ such that $\sum\limits_{i=1}^{\ws}\frac{1}{H(i)}\leq C$.
	Moreover, we obviously have $$\max\limits_{i\in E}(f_1(i)-\nu(i)t)\leq\max\limits_{i\in \mathbb{N}}(f_1(i)-\nu(i)t)$$, which leads to
	\begin{equation*}
		\sum\limits_{i=1}^{\ws}M_{m,i}\log\left(1-e^{-\frac{W_{m,i}}{\omega(m)}t\sum\limits_{j=1}^{p}g_j(m)}\right) \geq -2Ce^{2g_1(m)\max\limits_{i\in\mathbb{N}}(f_1(i)-\nu(i)t)}.
	\end{equation*}
	Then,
	\begin{eqnarray*}
&&\int_{t^\ast(f_1,\nu)}^{\infty}\left[1-\exp\left(\sum\limits_{i=1}^{\ws}M_{m,i}\log\left(1-e^{-\frac{W_{m,i}}{\omega(m)}t\sum\limits_{j=1}^{p}g_j(m)}\right)\right)\right]dt\\
	 &&\leq
	 \int_{t^\ast(f_1,\nu)}^{\infty}\left[1-e^{-2Ce^{2g_1(m)\max\limits_{i\in\mathbb{N}}(f_1(i)-\nu(i)t)}}\right]dt
	 \\
	&&\leq
	2C\int_{t^\ast(f_1,\nu)}^{\infty}e^{2g_1(m)\max\limits_{i\in\mathbb{N}}(f_1(i)-\nu(i)t)}dt.
	\end{eqnarray*}
	Choose $t^+>t^\ast(f_1,\nu)$, without any other assumption.
	As for all $t>t^\ast(f_1,\nu)$, $\max\limits_{i\in\mathbb{N}}(f_1(i)-\nu(i)t)<0$, and $g_1(m)\rightarrow+\infty$, we have $e^{2g_1(m)\max\limits_{i\in\mathbb{N}}(f_1(i)-\nu(i)t)}\underset{m\rightarrow\infty}{\rightarrow}0$.
	Then
	\begin{equation*}	
	\int_{t^\ast(f_1,\nu)}^{t^+}e^{2g_1(m)\max\limits_{i\in\mathbb{N}}(f_1(i)-\nu(i)t)}dt\underset{m\rightarrow\infty}{\rightarrow}0.
	\end{equation*}
	Besides, for all $t\geq t^+$, we have $f_1(i)-\nu(i)t\leq f_1(i)\frac{t}{t^+}-\nu(i)t$, hence
	\begin{equation*}
		\max\limits_{i\in E}(f_1(i)-\nu(i)t)\leq \max\limits_{i\in E}(f_1(i)\frac{t}{t^+}-\nu(i)t) = \frac{t}{t^+}\max\limits_{i\in E}(f_1(i)-\nu(i)t^+).
	\end{equation*}
	From Lemma \ref{lemfg} and \textbf{H3}, this last maximum, denoted $-\gamma$, is actually reached and we have $-\gamma=\max\limits_{i\in\mathbb{N}}(f_1(i)-\nu(i)t^+)<0$. Then,
	\begin{eqnarray*}
		\int_{t^+}^{\infty}e^{2g_1(m)\max\limits_{i\in\mathbb{N}}(f_1(i)-\nu(i)t)}dt &\leq& \int_{t^+}^{\infty}e^{-2\gamma g_1(m)\frac{t}{t^+}}dt\\
		&=& \frac{e^{-2\gamma g_1(m)}}{2\gamma g_1(m)}t^+\underset{m\rightarrow\infty}{\rightarrow}0
	\end{eqnarray*}
	and finally,
	\begin{equation}\label{eq2}	\int_{t^\ast(f_1,\nu)}^{\infty}\left[1-\prod\limits_{i=1}^{\ws}\left(1-e^{-\frac{W_{m,i}}{\omega(m)}t\sum\limits_{j=1}^{p}g_j(m)}\right)^{M_{m,i}}\right]dt \underset{m\rightarrow\infty}{\rightarrow}0.
	\end{equation}
	$\bullet$ Equations (\ref{eq1}) and (\ref{eq2}) lead to
	\begin{equation}\label{eq3}	\int_{0}^{\infty}\left[1-\prod\limits_{i=1}^{\ws}\left(1-e^{-\frac{W_{m,i}}{\omega(m)}t\sum\limits_{j=1}^{p}g_j(m)}\right)^{M_{m,i}}\right]dt \underset{m\rightarrow\infty}{\rightarrow}t^\ast(f_1,\nu).
	\end{equation}
	And finally, using \textbf{H1} and equation (\ref{eq3}),
	\begin{equation*}
		E[C_m] \sim t^\ast(f_1,\nu)\mu_m\sum\limits_{j=1}^{p}g_j(m) \sim t^\ast(f_1,\nu)\mu_m g_1(m).
	\end{equation*}
\hfill $\Box$

\end{document}